# Fundamental limits and near-optimal design of graphene modulators and non-reciprocal devices

Michele Tamagnone[1], Arya Fallahi[2], Juan R. Mosig[3], Julien Perruisseau-Carrier[1*]

[1] Adaptive MicroNano Wave Systems, Ecole Polytechnique Fédérale de Lausanne (EPFL), 1015 Lausanne, Switzerland

[2] DESY-Center for Free-Electron Laser Science (CFEL), Notkestrasse 85, D-22607 Hamburg, Germany

[3] Laboratory of Electromagnetics and Acoustics (LEMA), Ecole Polytechnique Fédérale de Lausanne (EPFL), 1015 Lausanne, Switzerland

**\*Corresponding Author:** julien.perruisseau-carrier@epfl.ch

**The potential of graphene for use in photonic applications was evidenced by recent demonstrations of modulators, polarisation rotators, and isolators. These promising yet preliminary results raise crucial questions: what is the optimal performance achievable by more complex designs using multilayer structures, graphene patterning, metal additions, or a combination of these approaches, and how can this optimum design be achieved in practice? Today, the complexity of the problem, which is magnified by the variability in graphene parameters, leaves the design of these new devices to time-consuming and suboptimal trial-and-error procedures. We address this issue by first demonstrating that the relevant figures of merit for the devices above are subject to absolute theoretical upper bounds. Strikingly these limits are related only to the conductivity tensor of graphene; thus, we can provide essential roadmap information such as the best possible device performance versus wavelength and graphene quality. Second, based on the theory developed, physical insight, and detailed simulations, we demonstrate how structures closely approaching these fundamental limits can be designed, demonstrating the possibility of significant improvement. These results are believed to be of paramount importance for the design of graphene-based modulators, rotators, and isolators and are also directly applicable to other 2-dimensional materials.**

In recent years, there has been a surge of interest in graphene-based devices for various photonic and optoelectronic applications[1-15]. Among other very promising passive devices, including amplitude modulators[5, 16-23], Faraday and Kerr polarisation rotators[24-26], and non-reciprocal isolators[27, 28] have been demonstrated. These essential photonic blocks might constitute some of the most important applications of graphene; thus, there is a significant and growing effort to transform the above-mentioned initial demonstrations into real capabilities. However, this prospect still requires drastic performance improvements. For this purpose, researchers have now started to develop graphene-based devices with multiple degrees of freedom by considering several technological solutions (see Figure 1e), including multilayer structures[22, 29, 30], graphene patterning[7, 22, 25, 31], substrate engineering, and combining metal and graphene[13, 14, 18, 21, 29, 32].

Except for very particular cases, structures as complex as those listed above cannot be modelled analytically; thus, the analysis of a single design variation entails intensive numerical electromagnetic simulations. Moreover, the best performance will typically be achieved by combining several of the above approaches, resulting in a very large optimisation space. Consequently, today, the design process is left to trial-and-error optimisation procedures that are based on heavy electromagnetic simulations; such procedures are not only time consuming but are also bound to yield sub-optimal performance or unnecessarily complex structures. Furthermore, the high sensitivity of the conductivity of graphene to fabrication or applied bias fields is a graphene-specific issue: given the lack of analytical solutions available, it is very impractical to distinguish potential performance improvements that are due to higher graphene quality from those due to improved device design.

Our results constitute a disruptive and far-reaching solution to the problems above. First, we demonstrate that there are absolute theoretical limits to the performance of graphene-based modulators, non-reciprocal polarisation rotators, and isolators. Remarkably, these upper bounds depend only on the conductivity of the graphene used to implement the devices. Thus, the minimum graphene quality, doping, and bias fields required to enable a given desired performance can be determined prior to any design or simulation. Similarly, for each application of interest, we calculate the best possible performance achievable as a function of frequency or any parameter influencing graphene conductivity, providing crucial 'roadmap' information for graphene photonics.

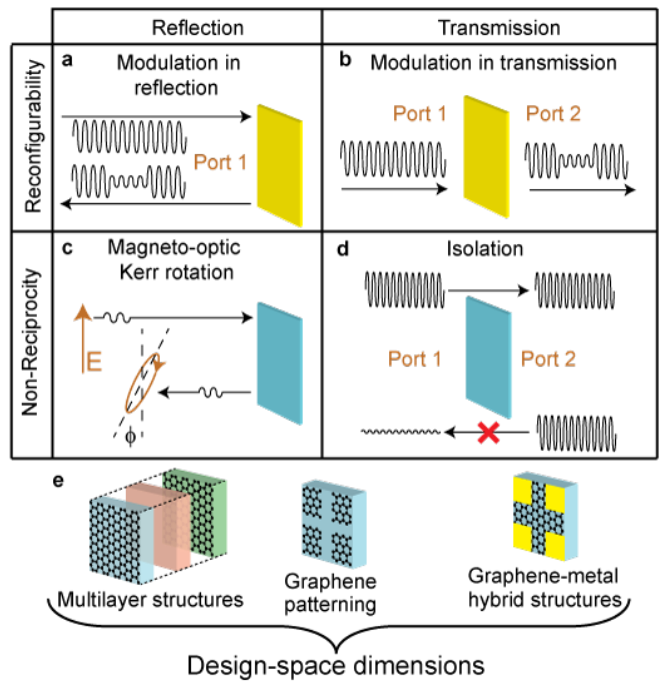

**Figure 1 | Graphene-based capabilities used for a detailed illustration of the method and 'design space' dimensions.** The device capabilities are classified according to the direction (reflection or transmission) and framework (non-reciprocity or reconfigurability). **a,** amplitude modulation in reflection. **b,** amplitude modulation in transmission. **c,** magneto-optical Kerr rotation. **d,** isolation. **e,** main independent degrees of freedom for the design of the devices.

The above fundamental results take on their full practical relevance only if one can design devices whose performance closely approaches the theoretical limit. We demonstrate that such devices can be designed on the basis of a large set of electromagnetic simulations, which also validate our theoretical results. Furthermore, based on the physical insight and concrete examples, we provide guidelines for designing low-complexity devices with quasi-optimal performance.

The results are based on rigorous theory and are applicable to virtually all graphene-based reconfigurable or non-reciprocal photonic devices, based on either guided or free-space waves. Additionally, the methodology can readily be applied to other 2D materials, such as $MoS_2$, or to 2D electronic gases (2DEGs) in semiconductors. For illustration purposes, four representative applications of significant recent interest are addressed in detail (Figure 1a-d), including amplitude modulation in reflection and transmission, magneto-optic Kerr rotation, and non-reciprocal isolation based on Faraday rotation.

## Electro-optical graphene amplitude modulators

Electro-optic modulation is the most studied application based on the dynamic reconfiguration of graphene conductivity[5, 16, 17, 19-22, 33-35]. The practical feasibility of these devices has been verified at different frequencies, ranging from infrared[5] to kHz[35]. In particular, graphene demonstrates a remarkable potential for modulation at THz frequencies, where alternative technologies exhibit significant limitations[16].

Graphene modulators, either in guided-wave systems[5] or as metasurfaces for free space beams[16, 21], can be rigorously described using scattering matrix formalism[36], which relates incident waves $\boldsymbol{a}$ to reflected waves $\boldsymbol{b}$ with $\boldsymbol{b} = \underline{S}\boldsymbol{a}$. The scattering matrix of the modulator takes two distinct values, $\underline{S}_A$ and $\underline{S}_B$, for the two scalar conductivities $\underline{\sigma}_A$ and $\underline{\sigma}_B$ of graphene, obtained by applying different electrostatic bias fields. Here, we consider amplitude modulators both in reflection (e.g., ref. 32) and in transmission (e.g., ref. 16). A modulator in reflection is a one-port device (Figure 1a) whose scattering matrix is a scalar with values $\underline{S}_A = \Gamma_A$ and $\underline{S}_B = \Gamma_B$. In contrast, for modulators in transmission, the input and output signals flow through different ports, and the scattering matrices are 2×2 and symmetrical:

$$\underline{S}_A = \begin{pmatrix} \Gamma_{1A} & T_A \\ T_A & \Gamma_{2A} \end{pmatrix} \quad ; \quad \underline{S}_B = \begin{pmatrix} \Gamma_{1B} & T_B \\ T_B & \Gamma_{2B} \end{pmatrix} \tag{1}$$

where $T_A$ and $T_B$ represent the transmission coefficients responsible for the modulation and the other terms represent undesired reflections.

Applying the mathematical formulation described in the Supplementary Methods to the scattering matrices defined above, the following inequalities are obtained:

$$\gamma_{modR}(|\Gamma_A|, |\Gamma_B|) \triangleq \frac{(|\Gamma_A| - |\Gamma_B|)^2}{(1 - |\Gamma_A|^2)(1 - |\Gamma_B|^2)} \leq \gamma_M \tag{2}$$

$$\gamma_{modT}(|T_A|, |T_B|) \triangleq \frac{(|T_A| - |T_B|)^2}{(1 - |T_A|^2)(1 - |T_B|^2)} \leq \gamma_M \tag{3}$$

where $\gamma_M$ is given by

$$\gamma_M = \frac{|\sigma_A - \sigma_B|^2}{4Re(\sigma_A)Re(\sigma_B)} \tag{4}$$

Inequalities (2) and (3) are hereafter referred to as the *amplitude modulation inequalities*. The left term $\gamma_{modR/T}$ is only a function of the two reflection/transmission coefficients, whereas the term on the right, $\gamma_M$, is a positive real scalar that depends only on the graphene conductivities $\underline{\sigma}_A$ and $\underline{\sigma}_B$, as shown in (4).

Helpful intuition regarding the inequality introduced is obtained by considering the Cartesian plane $(|\Gamma_A|, |\Gamma_B|)$, as depicted in Figure 2a. For brevity, only the case of reflection is considered here, but the illustration applies equally to transmission modulators due to the similarity of (2) and (3). Inequality (2) states that there are some areas in the space of $(|\Gamma_A|, |\Gamma_B|)$ that are strictly 'forbidden;' these regions are highlighted in yellow in Figure 2a. The boundaries of these forbidden regions are determined by $\gamma_M$ and thus solely by graphene conductivity. In this plot, an ideal modulator corresponds to the red squares, i.e., $|\Gamma_A| = 1$ and $|\Gamma_B| = 0$, or vice versa. However, (2) readily shows that such an ideal modulator is not practically realisable. For example, if the modulator is designed to achieve perfect absorption in its 'off' state, such that $|\Gamma_B| = 0$, then $|\Gamma_A|$ cannot exceed the value shown by the red circle in Figure 2a.

A more practical view is obtained by representing the same data in the Cartesian plane defined by insertion loss and modulation depth (Figure 2b). This representation is accomplished by assuming, without loss of generality, that $|\Gamma_A| \geq |\Gamma_B|$ and by defining the modulation depth as $h = (|\Gamma_A| - |\Gamma_B|)/(|\Gamma_A| + |\Gamma_B|)$ and the insertion loss as $|\Gamma_A|$. In this case, (2) writes:

$$\gamma_{modR}(|\Gamma_A|, h) \triangleq \frac{(2h|\Gamma_A|)^2}{(1 - |\Gamma_A|^2)((1+h)^2 - |\Gamma_A|^2(1-h)^2)} \leq \gamma_M \qquad (5)$$

For example, the best possible modulator with 100% modulation depth will have a loss of $|\Gamma_A| = \sqrt{\gamma_M/(1 + \gamma_M)}$, which is obtained by introducing $h = 1$ in (5). This case is represented by the red circle in Figure 2b, which corresponds exactly to the red circle in the alternative representation in Figure 2a.

In summary, the theory indicates that a given target value for modulation depth cannot be reached without a minimum value for insertion loss. The general trend is intuitive and related to the loss in graphene. More precisely, a high modulation originates from a strong interaction of the fields with graphene, which in turn increases the loss. However, here, we rigorously demonstrate that although arbitrarily complex designs might potentially allow the modulation depth to be increased without limit, such unbounded increases will *always* come at the cost of a minimum amount of loss. Importantly, this minimum amount of loss is a known function of the parameters of graphene alone, which allows us to fully utilise the developed theory for practical purposes, as detailed below.

The bound $\gamma_M$ is a function of graphene conductivity (9), and Figure 2c presents the frontiers corresponding to different values of $\gamma_M$. Modulators with different amplitudes that are available in the literature are also reported on the graph, where possible with their corresponding frontiers. The performance achieved in these initial concept demonstrations is typically significantly below the theoretical upper bound. This result suggests the important potential for improvement if devices approaching the theoretical limit can indeed be designed in practice.

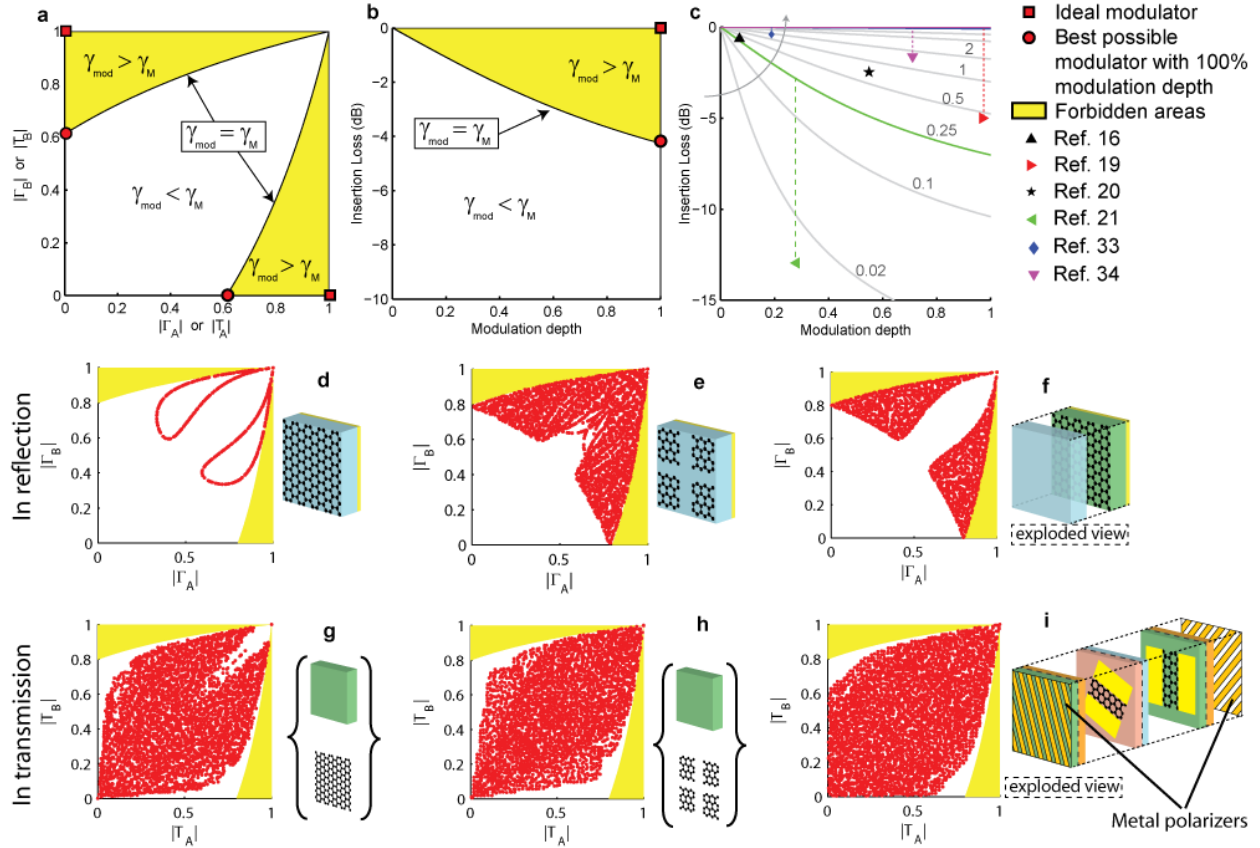

**Figure 2 | Performances of electro-optical modulators. a,** graphical representation of the amplitude modulation inequality in the Cartesian plane ($|\Gamma_A|, |\Gamma_B|$) [here, $\gamma_M = 0.6$ is used as an example, see (4)]. The squares represent ideal modulators, and the circles denote the best possible modulators with 100% modulation depth. Forbidden areas (yellow) are delimited by the boundary curve, where $\gamma_{mod} = \gamma_M$. **b,** same as **a** but using the insertion loss and modulation depth coordinates. **c,** upper bounds for different values of $\gamma_M$. The available designs in the literature are represented by coloured symbols, and where possible, the corresponding bound is represented using the same colour. **d-f,** simulations of randomly generated reflection modulators. Each red point represents a single simulated device. The frequency considered is 1 THz, and the graphene parameters are $T = 300$ K, $\mu_{cA} = 0.1$ eV, $\mu_{cB} = 0.8$ eV, and $\tau = 66\ fs$ (leading to $\gamma_M = 1.76$). A graphene layer on a back metallised dielectric layer (**d**) can reach optimal performance in a limited range. However, if graphene is patterned in a periodic square array (**e**) or if an additional dielectric layer is added (**f**), optimal performances can be reached along the entire boundary curve, including the best possible reflection modulation with 100% modulation depth. **g-i,** random simulations of different device topologies for transmission modulation. **g** represents random sequences of graphene sheets and dielectric layers. **h** represents random sequences of patterned graphene and dielectric layers. **i** shows an example of a complex structure employing polarisers and hybrid graphene metal structures showing near-optimal performances. Structures **d-h** are dual polarised, whereas the polarisers in **i** restrict the operation to single linear polarisation (with a 90° polarisation twist).

To provide our first answers to this question and subsequently to validate the theoretical prediction, as explained later, we simulated a very large number of *randomly generated* modulators. All modulators use graphene of the same conductivities and use the design degrees of freedom depicted in the insets of Figure 1e. The simulations are carried out using a homemade solver based on the periodic method of moments in the spectral domain. Compared to general-purpose commercial software, this tool not only allows significantly faster computations but also allows the magnetostatic bias to be considered, as needed for Kerr rotators later in this work (further details on the numerical tool and the randomly generated modulators are provided in Methods and Supplementary Methods). Graphene conductivity is evaluated using Kubo formalism[37] for the parameters reported in the caption of Figure 2, leading to $\gamma_M \cong 1.76$ and the corresponding theoretical upper bound plotted in Figure 2(d-e-f). The red dots in the figures report the computed performance of each randomly generated modulator.

The simplest possible reflection modulator, namely, uniform graphene over a metallised substrate (Figure 2d), can only achieve near-optimal performance for small modulation depths. It is also observed that in this case, the random simulations form a 1D locus. The same behaviour is observed by employing a multilayer dielectric stack between graphene and reflector. The reason is that the admittance seen at the input of the modulator, which uniquely determines the complex reflection coefficient, is the sum of graphene conductivity and the admittance seen at the input of the stack. The latter is always an imaginary number; thus, varying the permittivity, thickness, or number of layers provides only a single equivalent degree of freedom, which is the cause of the 1D locus.

Therefore, the logical next step is to consider the simplest modulators that provide additional design flexibility, namely, graphene patterning (Figure 2e) or the use of an additional dielectric layer before graphene (Figure 2f). The distribution of the computed results (red dots) provides evidence that these minimal increments in device complexity allow an almost arbitrary approach to the absolute theoretical upper bound, which is of considerable practical importance. The numerical results presented correspond to the conductivity of the given selected graphene, but the conclusions apply to any conductivity. Owing to space considerations, the detailed parameters and performance of designs close to the optimal performance are provided in the Supplementary Methods. In the current implementation of the method, the results are bandwidth-agnostic; thus, in practice, different near-optimal solutions can be compared and selected according to the bandwidth requirement if needed.

The simulations provide convincing evidence of the bound validity. The frontiers of the performance clouds for the randomly generated modulators almost exactly match the theoretical upper bound. Importantly, many more simulations of complex randomly generated setups were carried out, including *combinations* of a multilayer substrate, patterned graphene, and the addition of metal. No single result was found to exceed the theoretical limit.

The optimal performances for modulators in transmission are more difficult to attain. Despite closely approaching the bound for most modulation depth, non-patterned and patterned multilayer structures are suboptimal when a 100%-modulation is desired (Figures 2g-h). The physical interpretation is that in contrast to modulators in reflection, a good high-transmission modulation state not only requires limiting loss in graphene but also must independently impedance-match the system. The operation of a polarisation twist upon transmission, which is achieved in Figure 2i by using highly anisotropic hybrid graphene metal patterns, offers a new degree of freedom. The hybrid metasurfaces modulate one polarisation, whereas the other polarisation is left largely unaffected, allowing for the creation of feedback in the structure. However the improvement is only useful for a 100% modulation depth and comes at increased complexity, hence a simpler unpatterned multilayer structures as in Figure 2g is in general the best solution (see the Supplementary Methods for detailed parameters of designs close to the optimal performance).

We have just confirmed that it is possible to design modulators whose performance closely approaches the theoretical limit. Therefore, because the upper bound itself only depends on the conductivity of graphene, we can predict a priori the performance that optimised graphene modulators will achieve as a function of frequency and all other parameters that influence graphene conductivity. This prediction will provide crucial 'roadmap' information regarding the applicability of graphene modulators at different wavelengths and the required graphene quality to achieve a given desired performance. Figure 3 presents the theoretical upper bound as a function of different parameters influencing the conductivity of graphene in the two states of the modulator, namely, $\gamma_M(f, T, \mu_{cA}, \mu_{cB}, \tau)$, leading to the following conclusions. First, the best modulators can be designed between 10 and 100 THz and, evidently, for larger dynamic variations of the chemical potential. The performance sharply decreases at shorter wavelengths due to the well-known universal conductivity of graphene at optical frequencies[38] $(\pi/2)\,e^2/h \cong 6.09 \cdot 10^{-5} S$ ( $16.4\ k\Omega$ ), effectively preventing any modulation; hence, $\gamma_M = 0$. The transition frequency depends on the largest chemical potential (Figure 3c) and is coarsely approximated by $\mu_{cB}/h$, corresponding to the emergence of interband transitions. As expected, temperature is only important for low values of the lower chemical potential, whereas the graphene relaxation time $\tau$ has a very strong impact on the ultimate performance because it directly affects the loss mechanism.

As a numerical example, consider the use of graphene for modulation at λ=1.55 μm at 300 K, assuming a moderate chemical potential variation between $\mu_{cA} = 0.3$ eV and $\mu_{cB} = 0.6$ eV. The minimum required $\tau$ to reach 100% amplitude modulation with less than 1 dB of insertion loss is 20 fs.

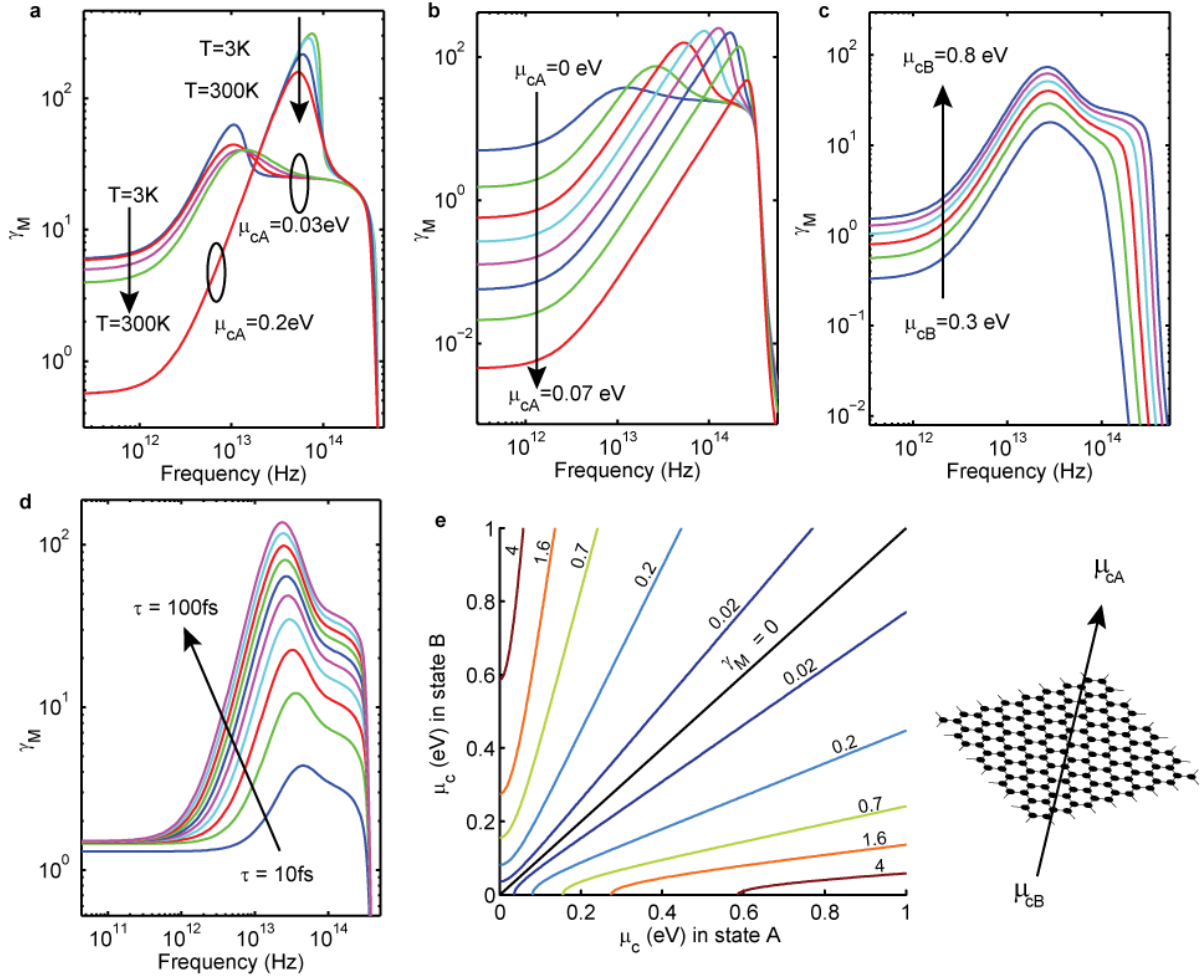

**Figure 3 | Theoretical upper bound ($\gamma_M$) to the performance of graphene modulators as a function of multiple parameters.** In all plots, the quantities that are not swept or otherwise specified have values of $f = 1$ THz, $T = 300$ K, $\mu_{cA} = 0.1$ eV, $\mu_{cB} = 0.8$ eV, $\tau = 66\ fs$. **a-d,** frequency dependence of $\gamma_M$ for several values of **a,** temperature, **b,** $\mu_{cA}$, **c,** $\mu_{cB}$, and **d,** $\tau$. **e,** parametric level curves of $\gamma_M$ for different values of $\mu_{cA}$ and $\mu_{cB}$.

# Non-reciprocal isolation

The second major class of graphene passive photonic devices is based on the presence of a magnetostatic field bias and the resulting off-diagonal terms in the conductivity tensor of graphene. This class includes Kerr and Faraday polarisation rotators, as well as isolators, which are non-reciprocal devices[24-30]. These devices are considered in the remainder of the paper. A similar procedure as that used in the case of the modulators can be followed concerning the mathematical derivation, its verification, and the practical exploitation of the results; thus, only the key methodology difference and practical results are considered in detail.

Non-reciprocal isolators display different transmission coefficients depending on the direction of wave propagation[27, 28] and can be completely described by a $2 \times 2$ scattering matrix:

$$S = \begin{pmatrix} S_{11} & S_{12} \\ S_{21} & S_{22} \end{pmatrix} \tag{6}$$

Assuming, without loss of generality, that $|S_{12}| \geq |S_{21}|$ and observing that here $S_A = S_B = S$, the following *isolation inequality* can be derived (see Supplementary Methods for the complete derivation):

$$\gamma_{isol}(|S_{12}|, |S_{21}|) \triangleq \frac{(|S_{12}| - |S_{21}|)^2}{(1 - |S_{21}|^2)(1 - |S_{12}|^2)} \leq \gamma_{NR} \tag{7}$$

where $\gamma_{NR}$ is given by

$$\gamma_{NR} = \frac{|\sigma_o|^2}{Re^2(\sigma_d) - Im^2(\sigma_o)} \text{ with } \underline{\sigma} = \begin{pmatrix} \sigma_d & \sigma_o \\ -\sigma_o & \sigma_d \end{pmatrix} \tag{8}$$

Notice that using the approximate Drude-Lorentz model[27], (8) becomes

$$\gamma_{NR} = (\omega_c \tau)^2 = (\mu B_0)^2 = (B_0 e \tau v_f^2)/\mu_c \tag{9}$$

where $\mu$ is the carrier mobility, $\omega_c$ is the cyclotron frequency, $v_f$ is the Fermi velocity, $e$ is the elementary charge, and $B_0$ is the magnetostatic biasing.

Similar to the case of the modulator (2), the left side of the inequality (7) is an intrinsic figure of merit of the isolator and is written as $\gamma_{isol}$. The right side is the *graphene non-reciprocity upper bound* $\gamma_{NR}$, which now takes the form in (8). Figures 4a and 4b represent the isolation inequality and the ideal and best possible optima. The isolation is defined as $I = |S_{12}|/|S_{21}|$. Figure 4c presents the variation in the best possible trade-off between isolation and insertion loss for different values of $\gamma_{NR}$. The minimum insertion loss for an isolator having perfect isolation ($I \to \infty$) is given by $|S_{12}| \leq \sqrt{\gamma_{NR}/(1 + \gamma_{NR})}$.

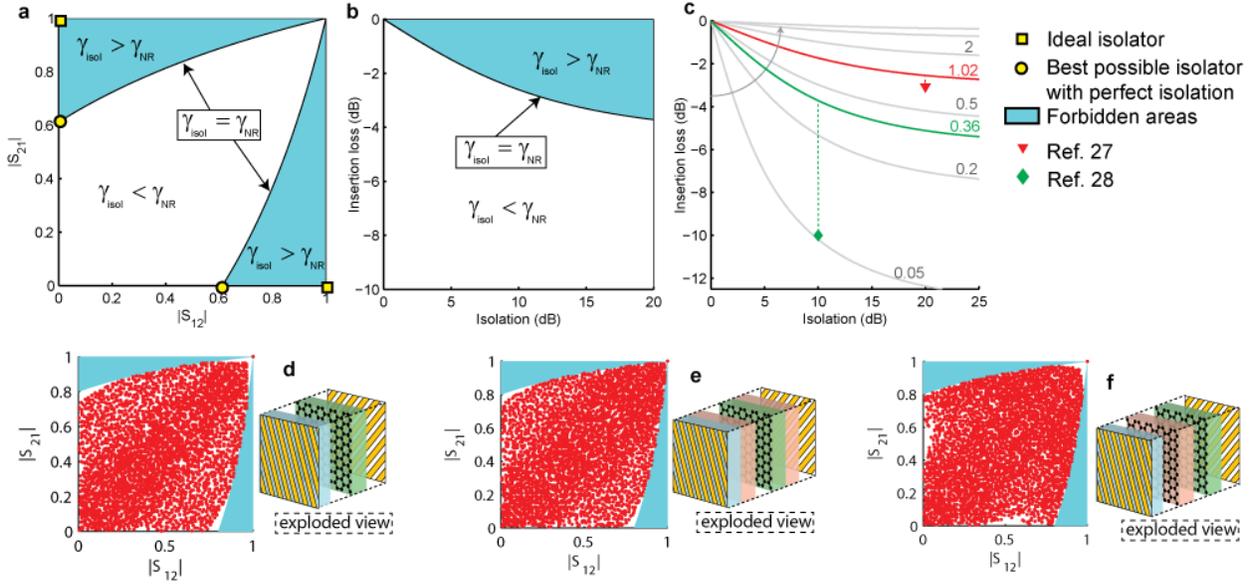

**Figure 4 | Performances of non-reciprocal isolators. a,** graphical representation of the isolation inequality on the Cartesian plane ($|S_{12}|, |S_{21}|$) (here, $\gamma_{NR} = 0.6$ is used as an example). The squares represent ideal isolators, and the circles denote the best possible isolators with perfect isolation. Forbidden areas (cyan) are delimited by the boundary $\gamma_{isol} = \gamma_{NR}$. **b,** same as **a** but using the insertion loss and the isolation of the isolator as coordinates. **c,** boundaries for different values of $\gamma_{NR}$. The available designs in the literature are represented by coloured symbols, and where possible, the corresponding boundary is represented using the same colour. **d-f,** random simulations for different device topologies. Each red point represents a single simulated device. The frequency considered is 1 THz, and the graphene parameters are $T = 3$ K, $\mu_c = 0.2$ eV, $B_0 = 4\ T$, and $\tau = 66\ fs$ ($\gamma_{NR} = 1.78$). All topologies use polarisers at both ends to convert Faraday rotation into isolation. **d** uses a graphene sheet enclosed by two dielectric layers. **e** uses two dielectrics on each side of the graphene sheet. **f** uses two graphene sheets and three dielectric layers in an alternating pattern.

Figures 4d to 4f illustrate that the best possible performance can be reached with simple planar structures encapsulated between two polarisers. In the simplest case of a graphene sheet placed between two dielectric slabs (Figure 4d), optimality is achieved everywhere except for a small degradation when perfect isolation is required ($I \to \infty$). Increasing the number of dielectric layers between graphene and polarisers does not solve this issue (Figure 4e). However, a structure comprised of three dielectric slabs and two graphene sheets in an alternating pattern can reach optimal performances at moderate complexity (Figure 4f). Unlike additional dielectric layers, a second graphene layer allows for decorrelating rotation and loss in the system, approaching the theoretical upper bound.

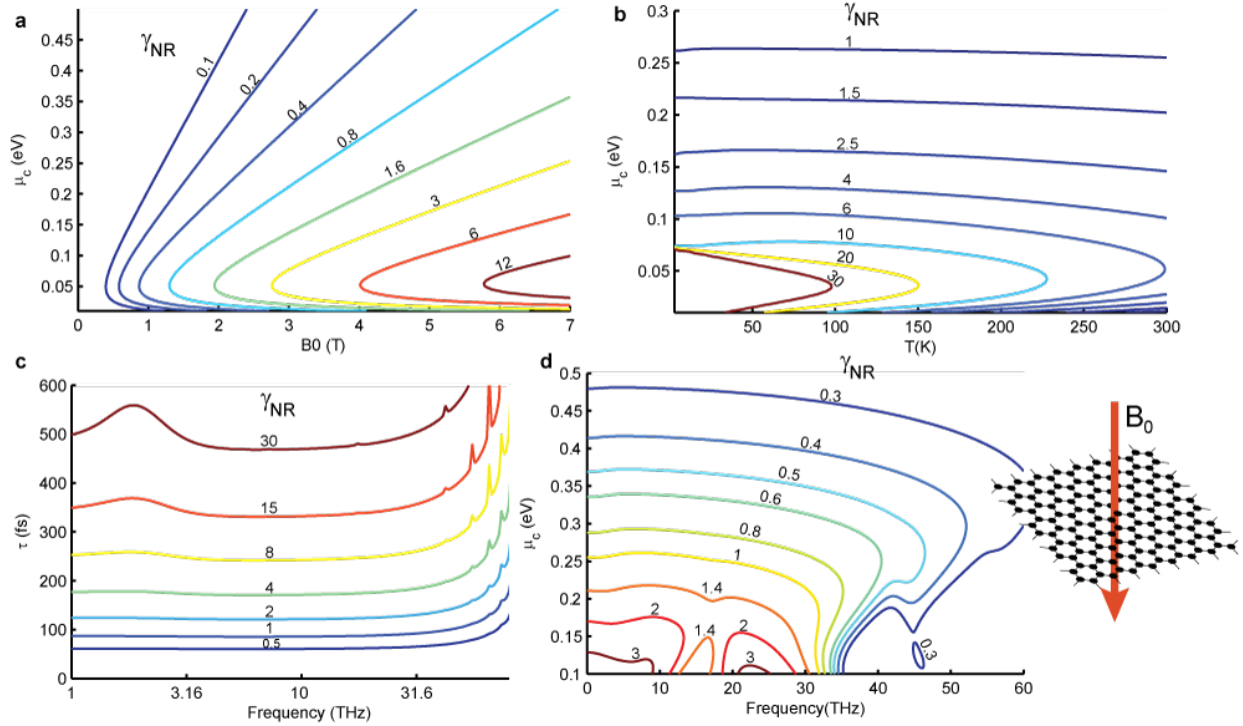

**Figure 5 | Theoretical upper bound ($\gamma_{NR}$) to the performance of graphene non-reciprocal devices as a function of multiple parameters.** In all plots, the quantities that are not swept have the values of $f = 1$ THz, $T = 300$ K, $\mu_c = 0.34$ eV, $B_0 = 4\,T$, and $\tau = 66\,fs$. Each contour is marked with the corresponding value of $\gamma_{NR}$. **a,** temperature vs. chemical potential sweep. **b,** bias magnetic field vs. chemical potential sweep. **c,** frequency vs. relaxation time sweep. **d,** frequency vs. chemical potential sweep.

As done earlier in the case modulators, we can now study the ultimate performance that optimally designed isolators can achieve as a function of graphene parameters via $\gamma_{NR}(f, T, \mu_c, B_0, \tau)$. Figure 5a illustrates that the optimal performance improves for larger magnetostatic biasing, as expected. A less obvious observation is the fact that isolators perform better at low $\mu_c$; this finding can also be inferred by the inspection of (9). This observation can be explained semiclassically by observing that the effective mass of the carriers decreases (or similarly, the mobility increases) for low $\mu_c$; thus, the bending of their trajectories due to the magnetic field increases (higher cyclotron frequency). Figure 5b illustrates that temperature is influential at low chemical potential, where the presence of thermal carriers of both polarities degrades the performance. High values of $\tau$, i.e., high graphene quality, can lead to very high $\gamma_{NR}$, as shown in Figure 5c. Figures 5c and 5d illustrate that performance is relatively frequency-invariant until the mid-infrared region, where the frequency drops due to interband transitions and to the universal optical conductivity of graphene[38]. The invariance toward low frequencies can be explained by the independence of $\gamma_{NR}$ on the imaginary part of $\sigma_d$ in (8). This behaviour contrasts with that of $\gamma_M$ in the modulation case, which degrades significantly toward low-terahertz and microwave frequencies (see Fig 3). This difference in behaviour indicates that unlike in modulators, plasmonic resonances are not instrumental to achieving high performance in graphene-based non-reciprocal polarisation rotators and isolators.

# Kerr rotation

An important effect associated with non-reciprocal surfaces is magneto-optical Kerr rotation, which is the rotation of the polarisation of a linear polarised wave upon reflection on a surface. The following *Kerr rotation inequality* is obtained (see the Supplementary Methods):

$$\gamma_{Kerr}(M,\varphi) \triangleq \frac{|2M \sin\varphi|^2}{(1-M^2)^2} \leq \gamma_{NR} \tag{10}$$

where $\varphi$ is the Kerr rotation angle, $M$ is the magnitude of the reflected linear polarisation, and $\gamma_{NR}$ is given by (8). $M$ represents the magnitude of the major ellipse axis when the polarisation of the reflected wave is elliptical. The inequality is graphically represented in Figure 6a-6c using the same conventions as in Figures 2 and 4. The 90° Kerr rotator is particularly important because it acts as a gyrator between vertically and horizontally polarised waves. The ideal and best possible Kerr rotators are identified by green markers in the figures.

Figure 6d presents the Kerr rotation of uniform graphene on a metal-backed substrate. Although optimal performance is obtained in a small region, the lack of degrees of freedom prevents optimal performance over the full optimal frontier, for the same reasons as explained in the case of modulators of the same structures. The insertion of a single superstrate of a dielectric provides an additional degree of freedom but has no particular effect on the performance (Figure 6e). However, two different superstrates allow for greatly enhanced Kerr rotations (Figure 6f), leading to optimal performances in a technologically simple structure. This enhancement is also in agreement with a similar effect reported for Faraday rotation[29,30]. Finally, a significant number of devices obtained by randomised combinations of the different strategies of Figure 6(d-f) were also simulated, with all of the devices satisfying the bound expressed in (14). The detailed parameters and results of the near-optimal designs are shown in the Supplementary Material.

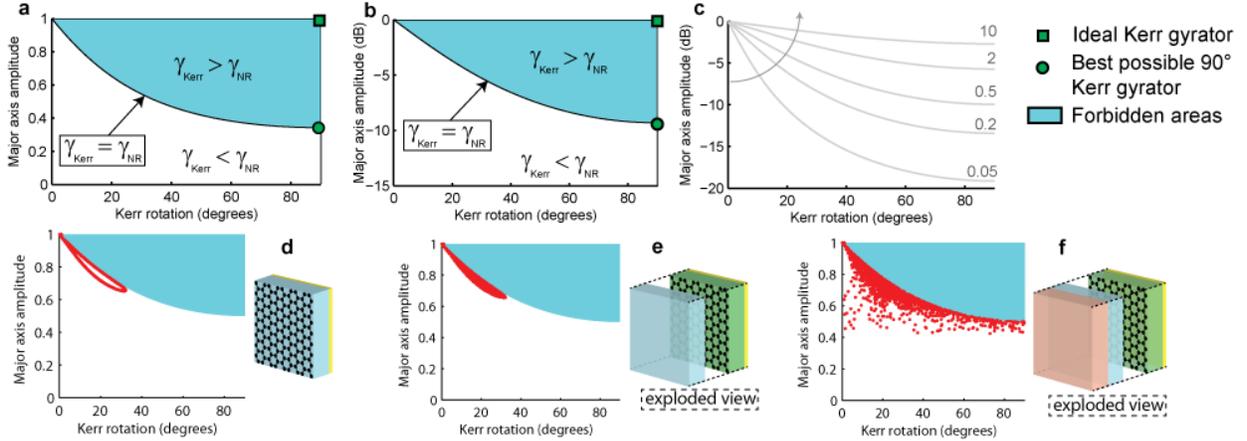

**Figure 6 | Performances of magneto-optical Kerr rotators. a,** graphical representation of the Kerr rotation inequality on the Cartesian plane $(M,\varphi)$ (here, $\gamma_{NR}=0.6$ is used as an example). The squares represent ideal 90° rotators, and the circles denote the best possible 90° rotators with perfect rotation. Forbidden areas (cyan) are delimited by the boundary curve, where $\gamma_{Kerr}=\gamma_{NR}$. **b,** same as **a** with the major axis expressed in dB. **c,** theoretical bound curve for different values of $\gamma_{NR}$. **d-f,** random simulations for different device topologies. Each red point represents a single simulated device. The frequency considered is 1 THz, and the graphene parameters are $T=3$ K, $\mu_c=0.2$ eV, $B_0=4$ T, and $\tau=66\ fs$ ($\gamma_{NR}=1.78$). The topology shown in **f** reaches optimal performance values for every rotation in the range 0°-90°. **d** is a single graphene layer on a back metallised substrate; in **e,** a superstrate is added; and in **f,** two superstrates are added, leading to the optimal performances.

# Conclusions

The performance of graphene-based modulators and non-reciprocal devices is bounded by absolute upper limits, which solely depend on the conductivity of the graphene employed. This relationship allows the ultimate performance that will be achieved by such devices to be predicted as a function of frequency and the other parameters that influence graphene conductivity. Simple technological implementations allow for very close approaches to the upper limit for the metasurface implementation of an amplitude modulator in reflection and transmission, as well as Kerr rotators and isolators. The observed influence of graphene parameters on the upper bounds, as well as the device topologies allowed to approach them, can be confirmed by physical insight. The developed theory applies to any passive linear structure that can be described in terms of a scattering matrix and can thus be extended to graphene guided devices and other 2DEGs. The developed methodology and practical results thereby obtained are believed to constitute an essential milestone toward the optimal operation of numerous future photonic devices.

# Methods

**Mathematical derivations**. The theoretical bounds are derived in two steps. First, a general scattering bound based on a generic scattering matrix is demonstrated using a proof inspired by a previous formulation for 3D materials used in guided devices[39]. Then, the relevant inequalities for actual graphene modulators, isolators, and Kerr rotators are obtained by algebraically manipulating the general bound. See the Supplementary Methods for more information and a complete proof of the different graphene device inequalities.

**Numerical simulations**. The simulations of biperiodic graphene structures have been performed using a periodic method of the moment electromagnetic solver. See the Supplementary Methods for more information.

# Acknowledgements

This work was supported by the Hasler Foundation (Project 11149) and by the Swiss National Science Foundation (SNSF) under grant 133583.